\documentclass[prd,nofootinbib,reprint,twocolumn,showpacs,superscriptaddress,floatfix]{revtex4-1}
\pdfoutput=1 
\usepackage{bm, color} 
\usepackage{amssymb,amsfonts,slashed,amsthm,amsmath, graphicx, soul, empheq, cancel}
\usepackage{hyperref}
\usepackage[caption=false]{subfig}

\begin{document}

\newcommand{\vev}[1]{ \left\langle {#1} \right\rangle }
\newcommand{\bra}[1]{ \langle {#1} | }
\newcommand{\ket}[1]{ | {#1} \rangle }
\newcommand{\eV}{ \ {\rm eV} }
\newcommand{\KeV}{ \ {\rm keV} }
\newcommand{\MeV}{\  {\rm MeV} }
\newcommand{\GeV}{\  {\rm GeV} }
\newcommand{\TeV}{\  {\rm TeV} }
\newcommand{\1}{\mbox{1}\hspace{-0.25em}\mbox{l}}
\newcommand{\Red}[1]{{\color{red} {#1}}}

\newcommand{\lmk}{\left(}  
\newcommand{\rmk}{\right)}
\newcommand{\lkk}{\left[}  
\newcommand{\rkk}{\right]}
\newcommand{\lhk}{\left \{ }  
\newcommand{\rhk}{\right \} }
\newcommand{\del}{\partial}  
\newcommand{\la}{\left\langle} 
\newcommand{\ra}{\right\rangle}
\newcommand{\half}{\frac{1}{2}}

\newcommand{\bea}{\begin{array}}
\newcommand{\eea}{\end{array}}
\newcommand{\beq}{\begin{eqnarray}}
\newcommand{\eeq}{\end{eqnarray}}
\newcommand{\eq}[1]{Eq.~(\ref{#1})}

\newcommand{\dd}{\mathrm{d}}
\newcommand{\Mpl}{M_{\rm Pl}}
\newcommand{\mg}{m_{3/2}}
\newcommand{\abs}[1]{\left\vert {#1} \right\vert}
\newcommand{\mphi}{m_{\phi}}
\newcommand{\Hz}{\ {\rm Hz}}
\newcommand{\for}{\quad \text{for }}
\newcommand{\Min}{\text{Min}}
\newcommand{\Max}{\text{Max}}
\newcommand{\Kahler}{K\"{a}hler }
\newcommand{\cphi}{\varphi}
\newcommand{\Tr}{\text{Tr}}
\newcommand{\diag}{{\rm diag}}

\newcommand{\SUf}{SU(3)_{\rm f}}
\newcommand{\Upq}{U(1)_{\rm PQ}}
\newcommand{\Zpq}{Z^{\rm PQ}_3}
\newcommand{\Cpq}{C_{\rm PQ}}
\newcommand{\ubar}{u^c}
\newcommand{\dbar}{d^c}
\newcommand{\ebar}{e^c}
\newcommand{\nubar}{\nu^c}
\newcommand{\Ndw}{N_{\rm DW}}
\newcommand{\Fpq}{F_{\rm PQ}}
\newcommand{\fpq}{v_{\rm PQ}}
\newcommand{\Br}{{\rm Br}}
\newcommand{\Lag}{\mathcal{L}}
\newcommand{\Lqcd}{\Lambda_{\rm QCD}}

\newcommand{\ji}{j_{\rm inf}} 
\newcommand{\jb}{j_{B-L}} 
\newcommand{\M}{M} 
\newcommand{\im}{{\rm Im} }
\newcommand{\re}{{\rm Re} }

\def\lrf#1#2{ \left(\frac{#1}{#2}\right)}
\def\lrfp#1#2#3{ \left(\frac{#1}{#2} \right)^{#3}}
\def\lrp#1#2{\left( #1 \right)^{#2}}
\def\REF#1{Ref.~\cite{#1}}
\def\SEC#1{Sec.~\ref{#1}}
\def\FIG#1{Fig.~\ref{#1}}
\def\EQ#1{Eq.~(\ref{#1})}
\def\EQS#1{Eqs.~(\ref{#1})}
\def\TEV#1{10^{#1}{\rm\,TeV}}
\def\GEV#1{10^{#1}{\rm\,GeV}}
\def\MEV#1{10^{#1}{\rm\,MeV}}
\def\KEV#1{10^{#1}{\rm\,keV}}
\def\blue#1{\textcolor{blue}{#1}}
\def\red#1{\textcolor{blue}{#1}}

\newcommand{\eff}{\Delta N_{\rm eff}}
\newcommand{\neff}{\Delta N_{\rm eff}}
\newcommand{\cc}{\Omega_\Lambda}
\newcommand{\Mpc}{\ {\rm Mpc}}
\newcommand{\Msolar}{M_\odot}


\title{
Small Instantons and the Post-Inflationary QCD Axion in a Special Product GUT
}
\author{Shihwen Hor$^{1,2}$, Yuichiro Nakai$^{1,2}$, Motoo Suzuki$^{3,4}$, and Junxuan Xu$^{1,2}$
\\*[10pt]
$^1${\it \normalsize Tsung-Dao Lee Institute, Shanghai Jiao Tong University, \\
No.~1 Lisuo Road, Pudong New Area, Shanghai, 201210, China} \\*[3pt]
$^2${\it \normalsize School of Physics and Astronomy, Shanghai Jiao Tong University, \\
800 Dongchuan Road, Shanghai, 200240, China} \\*[3pt]
$^3${\it \normalsize SISSA International School for Advanced Studies, \\
Via Bonomea 265, 34136, Trieste, Italy} \\
$^4${\it \normalsize INFN, Sezione di Trieste, Via Valerio 2, 34127, Italy}
}

\begin{abstract}

We present a new framework of grand unification that is equipped with an axion solution to the strong CP problem
without a domain wall problem when the Peccei–Quinn (PQ) symmetry is spontaneously broken after inflation.
Our grand unified theory (GUT) is based on a symmetry breaking pattern,
$SU(10) \times SU(5)_1 \rightarrow SU(5)_V \supset SU(3)_C \times SU(2)_L \times U(1)_Y$,
where $SU(5)_1$ and a special embedding of $SU(5)_2\subset SU(10)$ are broken to a diagonal subgroup $SU(5)_V$.
The model contains a vector-like pair of PQ-charged fermions that transform as (anti-)fundamental representations under $SU(10)$,
so that the domain wall number is one.
However, after the GUT symmetry breaking, the number of vector-like pairs of PQ-charged colored fermions is larger than one,
which seems to encounter the domain wall problem.
This apparent inconsistency is resolved by small instanton effects on the axion potential
which operate as a PQ-violating bias term and allow
the decay of domain walls.
We propose a domain-wall-free UV completion for an IR model where the domain wall number appears larger than one.
The model gives a prediction for a dark matter axion window,
which is different from that of the ordinary post-inflationary QCD axion with domain wall number one.

\end{abstract}

\maketitle

\section{Introduction
\label{introduction}}

The strong CP problem is a major unresolved puzzle in the Standard Model, as it requires an unnaturally small value for the CP-violating parameter in QCD.
A physical strong CP phase $\bar{\theta}$ would induce a neutron electric dipole moment (nEDM)
far larger than the experimental limits
\cite{Baker:2006ts,Pendlebury:2015lrz}.
The most compelling solution is the Peccei-Quinn (PQ) mechanism
\cite{Peccei:1977hh}, which introduces a global \( U(1)_{\text{PQ}} \) symmetry
that is spontaneously broken, leading to the emergence of a pseudo-Nambu-Goldstone boson called the axion
\cite{Weinberg:1977ma,Wilczek:1977pj}.
The axion dynamically relaxes $\bar{\theta}$ to zero by minimizing the QCD potential,
thus solving the strong CP problem naturally. Furthermore, if sufficiently light and weakly coupled,
the QCD axion can also serve as a viable dark matter (DM) candidate
\cite{Preskill:1982cy,Abbott:1982af,Dine:1982ah},
making it an attractive extension to the Standard Model.

The cosmological consequences of the axion primarily depend on two scenarios: (1) the \( U(1)_{\text{PQ}} \)  symmetry is broken during inflation and remains broken after reheating, or (2) it is restored during reheating and breaks spontaneously later.
In the former scenario, the axion field distribution is almost homogeneous across the Universe, and axion dark matter is produced through the misalignment mechanism~\cite{Preskill:1982cy,Abbott:1982af,Dine:1982ah}.
One drawback of this pre-inflationary scenario is 
the generation of large isocurvature fluctuations~\cite{Linde:1984ti,Seckel:1985tj,Lyth:1989pb,Turner:1990uz,Linde:1991km},
which are tightly constrained by the cosmic microwave background (CMB) measurements~\cite{Planck:2018jri},
requiring a (typically) fine-tuned low-scale inflation.

In the post-inflationary axion scenario, the axion field acquires spatial variations across the Universe.
This can lead to the formation of defects, such as global strings and domain walls.
If a model-dependent integer associated with the color anomaly, called the domain wall number \( N_{\text{DW}} \), is one,
domain walls form as disk-like structures attached to strings and eventually collapse due to their tension~\cite{Vilenkin:1982ks}.
The decay of the string-domain wall network produces a large number of axions, which dominate the axion DM abundance,
setting the PQ breaking scale around \( 10^{10} \) GeV~\cite{Hiramatsu:2012gg}.
On the other hand, if \( N_{\text{DW}} > 1 \), the domain walls form a stable network that eventually dominates the Universe, leading to a cosmological catastrophe known as the domain wall problem~\cite{Sikivie:1982qv}.
See refs.~\cite{Hagmann:1990mj,Battye:1993jv,Nagasawa:1994qu,Chang:1998tb,Yamaguchi:1998gx,Yamaguchi:1998gx,Hagmann:2000ja,Hiramatsu:2010yu,Hiramatsu:2012gg,Kawasaki:2014sqa,Fleury:2015aca,Klaer:2017qhr,Klaer:2017ond,Gorghetto:2018myk,Vaquero:2018tib,Kawasaki:2018bzv,Klaer:2019fxc,Buschmann:2019icd,Drew:2019mzc,Gorghetto:2020qws,Buschmann:2021sdq,Drew:2022iqz,Saikawa:2024bta,Kim:2024wku,Gorghetto:2024vnp,Benabou:2024msj} for numerical simulations of the evolution of strings and domain walls.

One potential solution to the domain wall problem is to introduce a small explicit breaking
of the \( U(1)_{\text{PQ}} \) symmetry, called a bias term,
which can lift the vacuum degeneracy and cause the domain walls to collapse before they dominate the Universe~\cite{Vilenkin:1981zs,Sikivie:1982qv,Gelmini:1988sf,Larsson:1996sp}.
This approach not only resolves the domain wall problem but also modifies the abundance of the axion DM,
as the early (late) decay of domain walls reduces (enhances) the axion relic density
compared to the conventional scenario with the domain wall number \( N_{\text{DW}} = 1 \).
However, in this solution, an ad hoc introduction of the bias term is unsatisfactory and raises a question about its origin.
To be worse, one has to adjust the bias term
so that the minimum of an extra axion potential generated is aligned with that of the QCD potential,
ensuring that the axion still solves the strong CP problem
\cite{Ghigna:1992iv,Barr:1992qq,Kamionkowski:1992mf,Dine:1992vx,Holman:1992us,Dobrescu:1996jp,Hiramatsu:2012sc}.

A possible origin of the bias term to address the domain wall problem is provided by small instanton effects on the axion potential~\cite{Holdom:1982ex,Holdom:1985vx,Flynn:1987rs,Poppitz:2002ac,Agrawal:2017ksf,Agrawal:2017evu,Csaki:2019vte,Gherghetta:2020keg,Fan:2021ntg,Cordova:2022ieu,Cordova:2023her,Csaki:2023ziz,Aoki:2024usv}.
Such instanton effects can arise from a hidden gauge sector beyond QCD.
That is, the infrared (IR) Standard Model gauge group is embedded into a larger ultraviolet (UV) gauge group,
whose natural candidate is offered by grand unified theory (GUT),
a compelling paradigm of physics beyond the Standard Model.
However, a naive embedding of the IR gauge group into a UV gauge group
such as $SU(3)_C \times SU(2)_L \times U(1)_Y \subset SU(5)$ does not work because
the resulting small instanton effects do not lift the vacuum degeneracy of the axion potential.

The solution to the issue of embedding is provided by the fact that 
simple Lie algebras possess not only regular subalgebras but also special subalgebras~\cite{Dynkin:1957ms,Dynkin:1957um}.
That is, while regular subalgebras are systematically obtained by removing nodes from Dynkin diagrams,
special subalgebras do not follow this scheme (see e.g.~\cite{Slansky:1981yr,Yamatsu:2015npn}).
To identify the IR gauge group as such a special subgroup of a UV gauge group
is essential to obtain small instanton effects resolving the vacuum degeneracy of the axion potential.
In the present paper, we propose a novel framework of grand unification
that is equipped with a post-inflationary axion solution to the strong CP problem
where the domain wall problem is addressed by a bias term induced by small instanton effects.
Our GUT model is based on a symmetry breaking pattern,
$SU(10) \times SU(5)_1 \rightarrow SU(5)_V \supset SU(3)_C \times SU(2)_L \times U(1)_Y$,
where $SU(5)_1$ and a special embedding of $SU(5)_2\subset SU(10)$ are broken to a diagonal subgroup $SU(5)_V$.
All the Standard Model matter fields are charged under the $SU(5)_1$ gauge group
where CP symmetry is spontaneously broken to generate the correct Cabibbo–Kobayashi–Maskawa (CKM) phase
without introducing a $\theta$ phase in $SU(5)_1$.\footnote{
Here, spontaneous CP violation itself does not solve the strong CP problem
because radiative corrections to the strong CP phase are not well-controlled
\cite{Dine:2015jga,Fujikura:2022sot}
(to make them under control requires additional structures such as supersymmetry
\cite{Fujikura:2022sot,Nakagawa:2024ddd} or extra dimension
\cite{Girmohanta:2022giy}).
}
The model contains a vector-like pair of PQ-charged fermions that transform as (anti-)fundamental representations under $SU(10)$,
so that the domain wall number associated with the $SU(10)$ is \( N_{\text{DW}} = 1 \).
After the GUT symmetry breaking, the number of vector-like pairs of PQ-charged colored fermions is larger than one,
due to the special embedding.
The apparent vacuum degeneracy is lifted by small instanton effects on the axion potential
that operates as a PQ-violating bias term, allowing the decay of domain walls.
Then, the model gives a prediction for a DM axion window,
which is different from that of the ordinary post-inflationary QCD axion with $N_{\rm DW} =1$.

The rest of the paper is organized as follows.
In section~\ref{specialGUT}, we first introduce the idea of special embedding and
then present our GUT model with the PQ mechanism.
Section~\ref{Axion_section} discusses the axion potential generated
by non-perturbative QCD effects and small instanton effects.
In section~\ref{cosmology_section}, we consider a post-inflationary axion scenario in our model
and find a viable parameter space.
Section~\ref{sec:Discussion} is devoted to conclusions and discussions.
Some model details are summarized in appendices.

\section{Special Product GUT}\label{specialGUT}

We here present our GUT model based on a symmetry breaking pattern,
$SU(10) \times SU(5)_1 \rightarrow SU(5)_V$,
where $SU(5)_1$ and a special embedding of $SU(5)_2\subset SU(10)$ are broken to a diagonal subgroup $SU(5)_V$.
Let us start with the introduction of special embedding.

\subsection{Special Embedding}

We focus on a gauge symmetry breaking $SU(2N)\to SU(N)$,
where $SU(2N)$ generators are denoted as $T_{\rm UV}^m$ $(m=1,...,(2N)^2-1)$
and those of $SU(N)$ are expressed as $ T_{\rm IR}^a$ $(a=1,...,N^2-1)$.
Each $ T_{\rm IR}^a$ is given by a linear combination of  $ T_{\rm UV}^m$,
\begin{align}
\label{eq:IR_O_UV}
    T_{\rm IR}^a=\mathcal{O}^{am} T_{\rm UV}^m \, , 
\end{align}
with coefficients $\mathcal{O}$.
For the ${\bf r}$ representation of $SU(N)$ that is embedded into the fundamental representation of $SU(2N)$,
the generators satisfy
\begin{align}
    \label{eq:trace_UV}
    &{\rm tr}(T_{\rm UV}^m T_{\rm UV}^n)=\frac{1}{2}\delta^{mn},\\[1ex]
    &{\rm tr}(T_{\rm IR}^a T_{\rm IR}^b)=T_{\rm IR}({\bf r})\delta^{ab}\ ,
    \label{eq:trace_IR}
\end{align}
where $T_{\rm IR}({\bf r})$ denotes the Dynkin index for the irreducible representation ${\bf r}$.
Substituting Eq.~\eqref{eq:IR_O_UV} into Eq.~\eqref{eq:trace_IR} and using Eq.~\eqref{eq:trace_UV}, we find
a condition on $\mathcal{O}$,
\begin{align}
\label{eq:uv_ir_o}
\mathcal{O}^{am}\mathcal{O}^{bn}\delta_{mn}=c \delta^{ab}\ ,
\end{align}
where $c$ is the ratio of Dynkin indices,
\begin{align}
    c\equiv \frac{T_{\rm IR}({\bf r})}{1/2}\ .
\end{align}
Special embedding, that we call, corresponds to the case with $c> 1$. 
Let us now consider a Weyl fermion $\psi$ that transforms as the fundamental representation of $SU(2N)$
but behaves as the ${\bf r}$ representation of the $SU(N)$ subgroup.
Its covariant derivative is expressed as
\begin{align}
    \mathcal{D}_\mu\psi&= \partial_\mu \psi -i g_{\rm UV} A^m_{{\rm UV},\mu} (T_{\rm UV}^m) \psi\\[1ex]
   &\supset \partial_\mu \psi -i g_{\rm IR} A^a_{{\rm IR},\mu} (T^a_{\rm IR})  \psi \, .
\end{align}
Here, $A^m_{{\rm UV},\mu}$ represents the $SU(2N)$ gauge field and $g_{\rm UV}$ is the corresponding gauge coupling,
while  $ A^a_{{\rm IR},\mu}$ denotes the $SU(N)$ gauge field and $g_{\rm IR}$ is its gauge coupling.
A part of the $SU(2N)$ gauge field can be expressed in terms of the  $SU(N)$ gauge field as
\begin{align}
    &   A^l_{{\rm UV},\mu} = \frac{g_{\rm IR}}{g_{\rm UV}}  A^a_{{\rm IR},\mu} (\mathcal{O})^{al} \, .
\end{align}
Requiring that the kinetic term of $ A^a_{{\rm IR},\mu}$ remains canonically normalized,
we find
\begin{align}
   g_{\rm IR}= g_{\rm UV}/\sqrt{c}\,\ .
\end{align}
In our GUT model, we consider the ${\bf r = 10}$ representation of $SU(5) \subset SU(10)$,
which leads to $c=3$.
A more explicit description focusing on this case is given in appendix~\ref{app:SU(5)}.

The theta term in the $SU(2N)$ gauge theory is related to that of the $SU(N)$ theory as
\begin{align}
    \int \frac{g_{\rm UV}^2}{8\pi^2}{\rm tr}(F_{\rm UV}\wedge F_{\rm UV})=
   \int \frac{c g_{\rm IR}^2}{8\pi^2}{\rm tr}(F_{\rm IR}\wedge F_{\rm IR})\ .
\end{align}
Here, $F_{\rm UV}, F_{\rm IR}$ denote the field strengths of the $SU(2N)$ and $SU(N)$ gauge fields, respectively.

\subsection{The $SU(10)\times SU(5)$ Model}

We consider a $SU(10) \times SU(5)_1$ gauge theory, where
the Standard Model (SM) quarks and leptons are embedded in the $\bar{\mathbf{5}}$ and ${\mathbf{10}}$ representations of $SU(5)_1$, as in the case of the conventional $SU(5)$ GUT paradigm. The SM Higgs field is also introduced within the ${\mathbf{5}}$ representation of $SU(5)_1$.\footnote{
To simply consider a special embedding of the ordinary $SU(5)$ into $SU(10)$, as proposed in ref.~\cite{Yamatsu:2017mei},
instead of a product GUT, encounters difficulties
of the appearance of extra fields charged under the SM gauge group, or the cancellation of gauge anomalies.
}
To achieve the symmetry breaking pattern $SU(10) \times SU(5)_1 \to SU(5)_V$, we introduce a Higgs field $\Phi$ transforming as the fundamental representation of $SU(10)$ and the $\overline{\mathbf{10}}$ representation of $SU(5)_1$:
\begin{align}
\Phi_a^{ij} : (\mathbf{10},~\overline{\mathbf{10}}) \ , 
\end{align}
where $a \, (=1-10)$ is the $SU(10)$ index, and $i, j \, (=1-5)$ are the $SU(5)_1$ indices.
As explained in appendix~\ref{app:SU(5)},
the form of the vacuum expectation value (VEV) of $\Phi$ that induces the desired breaking pattern
is described by the embedding of the $\overline{\mathbf{10}}$ representation of $SU(5)_1$
into the anti-fundamental representation of $SU(10)$.
By using the relation~\eqref{embeddingrelation}, the VEV takes the form,
\begin{align}
\langle \Phi \rangle = v \mathbf{1}_{10 \times 10} \ ,
\end{align}
where $\mathbf{1}_{10 \times 10}$ is the $10\times 10$ unit matrix, and $v$ is a parameter with mass dimension one.
Indeed, under infinitesimal transformations of $SU(10)$ and $SU(5)_1$, we have
\begin{align}
  \langle \Phi \rangle\to  (\mathbf{1}+ i\,\alpha_{10}^a T_{SU(10),{\bf 10}}^a-i\,\alpha_5^b T^b_{SU(5)_1,{\bf 10}})\langle \Phi \rangle\ ,
\end{align}
where $T_{SU(10),{\bf 10}}^a$ denote the generators for the fundamental representation of $SU(10)$
and $T^b_{SU(5)_1,{\bf 10}}$ are the generators for the {\bf 10} representation of $SU(5)_1$
embedded into the fundamental representation of $SU(10)$.
The remaining symmetry is given by the condition $\,\alpha_{10}^a T_{SU(10),{\bf 10}}^a-\,\alpha_5^b T^b_{SU(5)_1,{\bf 10}}=0$, corresponding to $SU(10)\times SU(5)_1\to SU(5)_V$.
For confirmation, appendix~\ref{gauge boson} discusses the gauge boson mass spectrum. 
A further breaking of $SU(5)_V \to [SU(3)_C \times SU(2)_L \times U(1)_Y] / \mathbf{Z}_6$ is induced by the VEV of an additional Higgs field in the ${\mathbf{24}}$ representation of $SU(5)_1$.
The matter content of the model is summarized in Tab.~\ref{tab:matter_contents}. The fermions ${\mathbf{10}}^{(1)}_f$ and $\bar{\mathbf{5}}^{(1)}_f$ ($f=1,2,3$) of $SU(5)_1$ correspond to three generations of the SM quarks and leptons. The Higgs fields are denoted by ${\mathbf{5}}^{(1)}_H$ and ${\mathbf{24}}^{(1)}_H$, with ${\mathbf{24}}^{(1)}_H$ potentially being either a real or complex scalar.

\begin{table}[t!]
    \centering
    \renewcommand{\arraystretch}{1.5}
    \begin{tabular}{|c|c|c|c|c|c|}
        \hline
        Field & Spin& $SU(10)$ & $SU(5)_1$ & $U(1)_{\rm PQ}$ & $U(1)_{\eta}$ \\ \hline
          $\Phi$ & 0 & $\mathbf{10}$ & $\overline{\mathbf{10}}$ & $0$ & 0  \\ \hline
        $\mathbf{10}_f^{(1)} (f=1-3)$ & 1/2 & $\mathbf{1}$ & $\mathbf{10}$ & $0$ & 0 \\ \hline
        $\bar{\mathbf{5}}_f^{(1)} (f=1-3)$ & 1/2 & $\mathbf{1}$ & $\bar{\mathbf{5}}$ & $0$ & 0  \\ \hline
        $\mathbf{5}_H^{(1)}$      & 0 & $\mathbf{1}$ & $\mathbf{5}$ & $0$ & 0 \\ \hline 
        $\mathbf{24}_H^{(1)}$      & 0 & $\mathbf{1}$ & $\mathbf{24}$ & $0$ & 0  \\ \hline
        \hline
         $\Psi_{99,f'}~(f'=1-4)$  & 1/2 & $\mathbf{99}$ & $\mathbf{1}$ & $0$ & 0 \\\hline
         $\Psi^{(1)}_{75,f'}~(f'=1-4)$  & 1/2 & $\mathbf{1}$ & $\mathbf{75}$ & 0 & 0
         \\\hline \hline
             $\psi$ & 1/2 & $\mathbf{10}$  & $\mathbf{1}$ & $+1$ & $-1$  \\ \hline
        $\bar\psi$ & 1/2 & $\mathbf{\overline{10}}$ & $\mathbf{1}$ & $0$ & $+1$ \\ \hline
            $\Phi_{\rm PQ}$ & 0 & $\mathbf{1}$ & $\mathbf{1}$ & $-1$ & 0 \\ \hline
        \hline
        $\eta_{\alpha} (a=1,2)$  & 0 & $\mathbf{1}$ & $\mathbf{1}$ & $0$ & $-1$ \\\hline
    \end{tabular}
    \vspace{3mm}
    \caption{The model's matter content.}
    \label{tab:matter_contents}
\end{table}

In order to address the strong CP problem, the model respects the $U(1)_{\rm PQ}$ symmetry
that is spontaneously broken by a complex scalar field $\Phi_{\rm PQ}$ and
contains a vector-like pair of $U(1)_{\rm PQ}$-charged fermions $\psi, \bar\psi$ that transform as
(anti-)fundamental representations under $SU(10)$,
so that the domain wall number \( N_{\text{DW}} = 1 \).
We call them as KSVZ fermions~\cite{Kim:1979if,Shifman:1979if}.
Their charge assignments are also summarized in Tab.~\ref{tab:matter_contents}.

In non-supersymmetric models,  a precise gauge coupling unification is not automatic.
To achieve unification at a high energy, one can introduce additional light fields that form incomplete GUT multiplets. Specifically, we consider to include sets of an $SU(2)_L$ triplet fermion and an $SU(3)_C$ octet fermion~\cite{Aizawa:2014iea}, where unification can be achieved without introducing fields carrying the $U(1)_Y$ charge.
Several approaches exist for embedding these extra fermions into GUT multiplets.
One option is to introduce the adjoint representations of $SU(5)_1$, while another possibility is to utilize the adjoint representation $\mathbf{99}$ of $SU(10)$.
For demonstration, we opt for the latter, introducing four sets of fermions, $\Psi_{99,f'}~(f'=1-4)$, in the $\mathbf{99}$ representation of $SU(10)$, where all components except the $SU(2)_L$ triplet and $SU(3)_C$ octet acquire masses near the Planck scale.
Concretely, under $SU(5) \subset SU(10)$, the $\mathbf{99}$ is decomposed  as
\begin{align}
    \mathbf{99}=\mathbf{75}\oplus \mathbf{24}\ .
\end{align}
To render the $\mathbf{75}$ components heavy, we introduce Weyl fermions $(\Psi^{(1)}_{75,f'})^{kl}_{ij}~(f'=1-4)$ in the $\mathbf{75}$ representation of $SU(5)_1$, which acquire masses near the Planck scale through
\begin{align}
\label{eq:99_75}
    \mathcal{L}
    \sim (\Psi_{99})^a_b \Phi_a^{ij}\Phi^{\dagger b}_{kl} 
    (\Psi^{(1)}_{75})^{kl}_{ij}\ ,
\end{align}
where the flavor indices are omitted, the Planck mass is set to $M_{\rm Pl}=1$, and $(\Psi^{(1)}_{75})^{kl}_{ij}$ satisfies the constraints~\cite{Georgi:1981vf},
\begin{align}
  &~(\Psi^{(1)}_{75})^{ij}_{kl}=-(\Psi^{(1)}_{75})^{ji}_{kl},~(\Psi^{(1)}_{75})^{ij}_{kl}=-(\Psi^{(1)}_{75})^{ij}_{lk},~
  (\Psi^{(1)}_{75})^{ij}_{ik}=0\ .
\end{align}
To ensure that the triplet and octet components remain light, a fine-tuning is required between the mass term of $\Psi_{99}$ and the following terms:
\begin{align}
    \mathcal{L}\sim \, &{\rm tr}(\Psi_{24} {\bf 24}_H^{(1)} \Psi_{24})
    +{\rm tr}(\Psi_{24} {\bf 24}_H^{(1)} \Psi_{24}  {\bf 24}_H^{(1)}) \nonumber \\
    &+{\rm tr}(\Psi_{24} {\bf 24}_H^{(1)}){\rm }\,
    {\rm tr}(\Psi_{24}  {\bf 24}_H^{(1)})  \ ,
\end{align}
where $(\Psi_{24})^i_j\equiv (\Psi_{99})^a_b\Phi_a^{k i}\Phi^{\dagger b}_{kj}$.
This fine-tuning issue will be discussed in section~\ref{sec:Discussion}.

\begin{figure}[!t]
    \centering
    \includegraphics[width=0.4\textwidth]{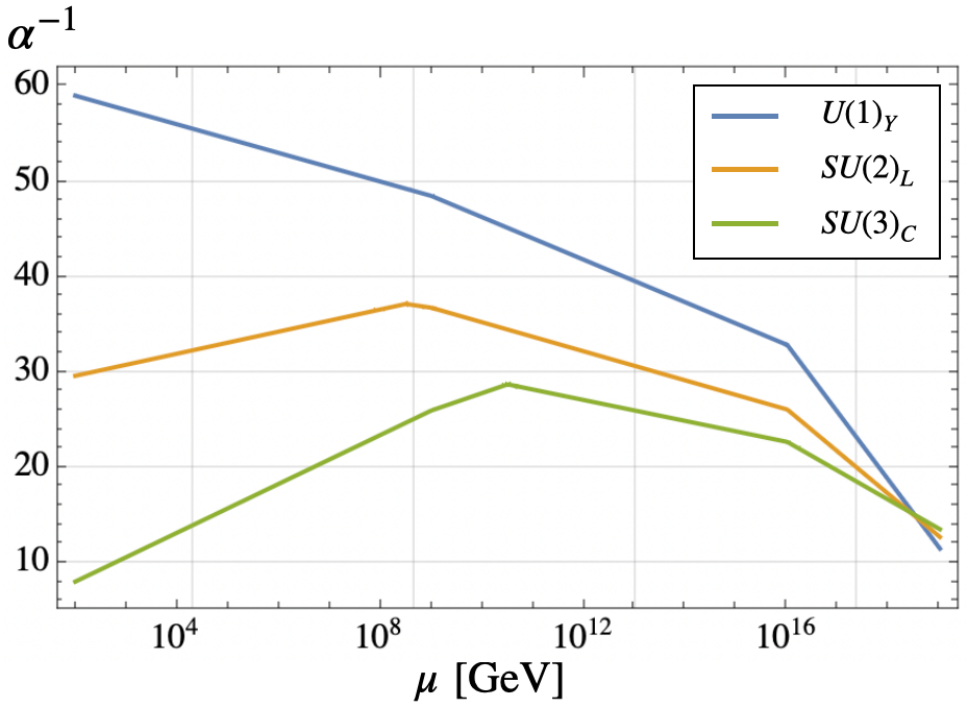} 
    \caption{One-loop renormalization group evolution of the inverse gauge couplings $\alpha^{-1} \equiv 4\pi/g^2$ of $U(1)_Y$, $SU(2)_L$ and $SU(3)_C$ (from top to bottom).}
    \label{fig:unification}  
\end{figure}

Fig.~\ref{fig:unification} shows the renormalization group evolution of the SM gauge couplings in the presence of four sets of light $SU(2)_L$ triplet and $SU(3)_C$ octet fermions, whose masses are taken as $10^{8.5}$\,GeV and $10^{10.5}$\,GeV, respectively, while the other components in the ${\bf 24}$ multiplets are as heavy as $10^{16}$\,GeV.
The KSVZ vector-like fermion mass is taken as $10^9$\,GeV.
The unification scale is found to be close to the Planck mass scale.
Above the unification scale, we assume the SM gauge couplings are unified to $SU(5)_V$ and soon above we have $SU(10)\times SU(5)_1$.

\subsection{
Spontaneous CP Violation
}

We assume that CP symmetry is present in the UV and consider its spontaneous violation in the IR.
As will be discussed in a later section, this structure allows us to align the minimum of the axion potential generated by small instanton effects,
which serves as a bias term to avoid the domain wall problem,
with that determined by non-perturbative QCD effects, thereby avoiding fine-tuning of complex phases.

To reproduce the CKM phase while maintaining an approximate alignment of phases, we introduce complex scalar fields $\eta_\alpha$ $(\alpha=1,2)$, which spontaneously breaks CP symmetry,
\begin{align}
   {\rm arg}( \langle \eta_\alpha\rangle )=\mathcal{O}(1)\ .
\end{align}
These scalar fields couple to the mixing term between the KSVZ fermion sector and the SM sector,
\begin{align}
\label{nelsonbarr}
  \mathcal{L}&\sim  
  \Phi_{\rm PQ} \psi \bar\psi
    +\sum_{f=1-3,\alpha=1,2}
    a^u_{\alpha f}
   \eta_\alpha
    \bar\psi^a  (\Phi)_a^{ij} {\bf 10}^{(1)}_{ij,f} \nonumber \\
   & +y_{ff'}^u{\bf 10}^{(1)}_{f}  {\bf 10}^{(1)}_{f'} {\bf 5}_H
    +y_{ff'}^d{\bf 10}^{(1)}_{f}  \bar {\bf 5}^{(1)}_{f'} {\bf 5}^\dagger_H\ ,
\end{align}
where all the coefficients $ a^u_{\alpha f}, y_{ff'}^u, y_{ff'}^d$ are real.
The first term represents the KSVZ fermion mass term, the second term introduces the mixing and CP violation,
and the third and fourth terms correspond to the quark and lepton Yukawa couplings.
The summation in the second term is explicitly written.

The present setup bears a similarity to the Nelson-Barr type mechanism~\cite{Nelson:1983zb,Barr:1984qx,Barr:1984fh}.
In the IR, the interactions \eqref{nelsonbarr} generate up-type quark mass terms:
\begin{align}
    \mathcal{L}
    &\sim
    (q_{uf}\,\ U\,\, Q_u )
  \mathcal{M}_u
     \begin{pmatrix}
     \bar u_{f'}\\
     \bar U\\
     \bar Q_u
     \end{pmatrix}\ ,  \\[1ex]
     \mathcal{M}_u&=  \begin{pmatrix}
     ( m_u)_{ff'} & 0 & A^*\\
    A^\dagger & v_{\rm PQ}  & 0\\
     0 & 0 &v_{\rm PQ}\\
     \end{pmatrix}\ , \label{uptypemass} \\[1.5ex]
     A^*&=\sum_{\alpha}a^u_{\alpha f} \eta_\alpha
      \ ,
\end{align}
where $\bar U,~Q_u$ denote the $SU(2)_L$ singlet and doublet up-type quarks in $\psi$, respectively,
while $U,~\bar Q_u$ are their vector-like pairs in $\bar \psi$,
and $\bar u_f,q_{uf}$ represent the $SU(2)_L$ singlet and doublet
up-type quarks of $\mathbf{10}^{(1)}_f$.
The $3\times 3$ mass matrix is defined as $(m_u)_{ff'}\equiv y^u_{ff'}v_{\rm SM}$
with the SM Higgs VEV $v_{\rm SM}$.
The down-type quark mass terms are also given by
\begin{align}
    \mathcal{L}
    &\sim
    (q_{d f}\,\, Q_d )
  \mathcal{M}_d
     \begin{pmatrix}
     \bar d_{f'}\\
     \bar Q_d
     \end{pmatrix}\ ,\\[1ex]
     \mathcal{M}_d&=  \begin{pmatrix}
     ( m_d)_{ff'} & A^*\\
       0 & v_{\rm PQ}  
     \end{pmatrix}
      \label{downtypemass}\ ,
\end{align}
where $\bar d_f,q_{df}$ represent the $SU(2)_L$ singlet and doublet
down-type quarks of $\mathbf{10}^{(1)}_f$, and $(m_d)_{ff'}\equiv y^d_{ff'}v_{\rm SM}$.
The low-energy effective $3\times 3$ matrices after integrating out the heavy quarks are computed in appendix~\ref{app:meff}
and given in Eqs.~\eqref{effective_up_mass}, \eqref{effective_down_mass}.
When $(a^u\langle\eta\rangle)_f\gtrsim v_{\rm PQ}$, an $\mathcal{O}(1)$ CKM phase is properly generated.

Since the determinant of each mass matrix in 
Eqs.~\eqref{uptypemass}, \eqref{downtypemass} is real,
the physical $\theta$-parameters of $SU(10)$, $SU(5)_1$, or $SU(3)_C$ vanish at the tree-level.
This will ensure an alignment of the axion potential minima from $SU(3)_C$ non-perturbative effects and the $SU(10)$ small instanton effect at least at the $\mathcal{O}(1)\%$ level due to a loop suppression.

Note that the Lagrangian~\eqref{nelsonbarr} respects an anomaly free $U(1)_\eta$ symmetry, under which $\eta_\alpha, \psi, \bar{\psi}$ transform with charges summarized in Tab.~\ref{tab:matter_contents}.
At the classical level, this symmetry can forbid dangerous terms that could induce sizable corrections to the physical $\theta$-parameter of 
$SU(3)_C$ denoted as $\bar \theta_c$,
\begin{align}
\label{eq:u1}
    \eta U\bar U, \quad \eta Q\bar Q, \quad HQ\bar u, \quad Hq\bar U\ ,
\end{align}
where $H$ denotes the SM Higgs field. 
However, radiative corrections can still generate 
corrections to $\bar \theta_c$,
which are estimated at the one-loop level as~\cite{Dine:2015jga}
\begin{align}
 \bar \theta_c\approx \frac{1}{16\pi^2 M_{\rm CP}^2}\Big| C_{\beta\gamma} a^u_{\alpha f} a^u_{\beta  f}
    \langle \eta_\alpha\rangle \langle \eta_\gamma\rangle^*\Big|\ ,
\end{align} 
where $M_{\rm CP}$ represents the mass of $\eta$'s, $C_{\beta\gamma}$ is the coefficient of the $|H|^2\eta^*_\beta \eta_\gamma$ interaction.
Two-loop corrections are given in similar forms, but receive an additional suppression by a loop factor, assuming $\mathcal{O}(1)$ coefficients for the self-interactions of $\eta$'s~\cite{Dine:2015jga}.
Further corrections to $\bar \theta_c$ may come from Planck-suppressed operators.
To prevent them, we impose $|\langle \eta_\alpha\rangle|\lesssim 10^{16}$\, GeV.
On the other hand, to avoid the CP domain wall problem~\cite{McNamara:2022lrw,Perez:2023zin}, we assume 
\begin{align}
  v_{\rm PQ} \ll  T_R\ll|\langle\eta_a\rangle|\ ,
\end{align}
where $T_R$ denotes the reheating temperature.

\section{Axion potential}\label{Axion_section}

We now estimate the axion potential generated by non-perturbative QCD effects and small instanton effects
in our $SU(10)\times SU(5)$ model with the PQ mechanism.

\subsection{QCD Effects}

Our GUT model is based on a $SU(10) \times SU(5)_1$ gauge theory.
In the UV, the theta-terms are given by
\begin{align}
\label{eq:theta_UV}
    \text{UV}: \int  \frac{\theta_{10}g^2_{10}}{8\pi^2}{\rm tr}(F_{10}\wedge F_{10})+
     \frac{\theta_{5}g_5^2}{8\pi^2}{\rm tr}(F_{5}\wedge F_{5})\ ,
\end{align}
where $F_{10}$, $F_5$ denote the field strengths of $SU(10)$ and $SU(5)_1$, respectively,
$g_{10}$, $g_5$ are the corresponding gauge couplings,
and $\theta_{10,5}$ are the theta-parameters of $SU(10)$ and $SU(5)_1$. 
After the breaking of $SU(10)\times SU(5)_1\to SU(5)_V$,
in the IR, the theta term turns into
\begin{align}
     \text{IR}: \int  \frac{\theta_c g^2}{8\pi^2}{\rm tr}(F\wedge F)\ ,
\end{align}
where $F$ denotes the field strength of the $SU(5)_V$ gauge field with gauge coupling.
\begin{align}
    \frac{1}{g^2}=\frac{3}{g_{10}^2}+\frac{1}{g_5^2}\ ,
\end{align}
Here, the factor 3 comes from our special embedding of $SU(5)_2 \subset SU(10)$.
The theta parameter of $SU(5)_V$ (and $SU(3)_C$) is given by $\theta_c= 3\theta_{10}+\theta_5$.%
\footnote{The physical theta angle of $SU(3)_C$ also includes the phases from the quark sector.}

The model contains a gauge singlet complex scalar field $\Phi_{\rm PQ}$ (PQ field) and a vector-like pair of KSVZ fermions $\psi,\bar\psi$ that transform as (anti-)fundamental representations of $SU(10)$. We assume the presence of a PQ symmetry,
under which these fields transform as
\begin{align}
    \Phi_{\rm PQ}\to e^{i\alpha}\Phi_{\rm PQ}\ , \quad \psi\to e^{-i\alpha}\psi, \quad \bar\psi\to \bar\psi\ .
\end{align}
The Lagrangian includes a PQ-invariant term,
\begin{align}
    \mathcal{L}\ni \Phi_{\rm PQ}\psi\bar\psi\ .
\end{align}
The PQ symmetry is spontaneously broken by a VEV of $\Phi_{\rm PQ}$,
giving a mass to the KSVZ fermions.
The phase component of $\Phi_{\rm PQ}$ corresponds to the axion,
\begin{align}
    \Phi_{\rm PQ}\sim v_{\rm PQ} e^{i\theta_a}\ ,
\end{align}
where the radial component is omitted for notational simplicity.
The KSVZ fermions transform as $\bf 10, \overline{10}$ representations under $SU(5)_V$,
and consequently, a vector-like pair of the $SU(2)_L$ doublet KSVZ (anti-)quarks as well as a pair of the $SU(2)_L$
singlet (anti-)quarks appear
after the GUT breaking of $SU(5)_V \rightarrow SU(3)_C \times SU(2)_L \times U(1)_Y$.
Hence, the axion coupling to the $SU(3)_C$ gauge field takes the form,
\begin{align}
\int (3\theta_a+\theta_c)\frac{g^2}{8\pi^2} {\rm tr}(G\wedge G)\ .
\end{align}
Here, $G$ denotes the field strength of the $SU(3)_C$ gauge field.
Then,
non-perturbative QCD effects generate the axion potential~\cite{GrillidiCortona:2015jxo},
\begin{align}
\label{QCDeffect}
    V_{\rm QCD}\approx
    -\frac{m_um_d}{(m_u+m_d)^2}m_\pi^2f_\pi^2 \cos\left(3\theta_a+\bar\theta_c\right)  ,
\end{align}
where $\bar{\theta}_c$ is the physical phase,
$m_\pi$, $f_\pi$ denote the pion mass and decay constant, $m_u$ and $m_d$ are the masses of the up and down quarks. As discussed in the previous section, $\bar\theta_c\ll 10^{-2}$ represents a phase contribution from radiative corrections with spontaneous violation of CP symmetry.
Note that the above axion potential seems to correspond to the case with the domain wall number $N_{\rm DW} =3$.

\subsection{Small Instanton Effects}

We now discuss the axion potential generated from a small instanton effect.
The model includes one flavor of KSVZ fermions in the (anti-)fundamental representations of $SU(10)$.
Additionally, we introduce four flavors of Weyl fermions in the $\mathbf{99}$ representation of $SU(10)$ and $\mathbf{75}$ representation of $SU(5)_1$ so that only $\mathbf{24}$ multiplets of $\mathbf{99}$ remain light compared to the Planck mass scale.
We then assume an approximate chiral symmetry,
\begin{align}
    \Psi_{99}\to \Psi_{99}e^{i\beta}, \quad \Psi_{75}^{(1)}\to \Psi_{75}^{(1)} e^{-i\beta}\ .
\end{align}
As a result, mass terms for $\Psi_{99}$ and $\Psi^{(1)}_{75}$ are suppressed by a small parameter $\kappa\ll 1$,
\begin{align}
\label{eq:kappa_99}
 \mathcal{L}
    \sim
\kappa^2 M (\Psi_{99})^a_b (\Psi_{99})^b_a
+
\kappa^{\dagger 2} M (\Psi^{(1)}_{75})^{kl}_{ij}
(\Psi^{(1)}_{75})^{ij}_{kl}\ .
\end{align}
Here, the dimensionless parameter $\kappa^2$ acts as a spurion for the chiral symmetry,
and the mass parameter $M \sim M_{\rm Pl}$ denotes the breaking scale of $SU(10)\times SU(5)_1\to SU(5)_V$.
Consequently, the $\mathbf{24}$ multiplet within $\mathbf{99}$ acquires a mass of $\mathcal{O}(\kappa^2 M)$.

\begin{figure}[t!]
    \centering
    \includegraphics[width=0.4\textwidth]{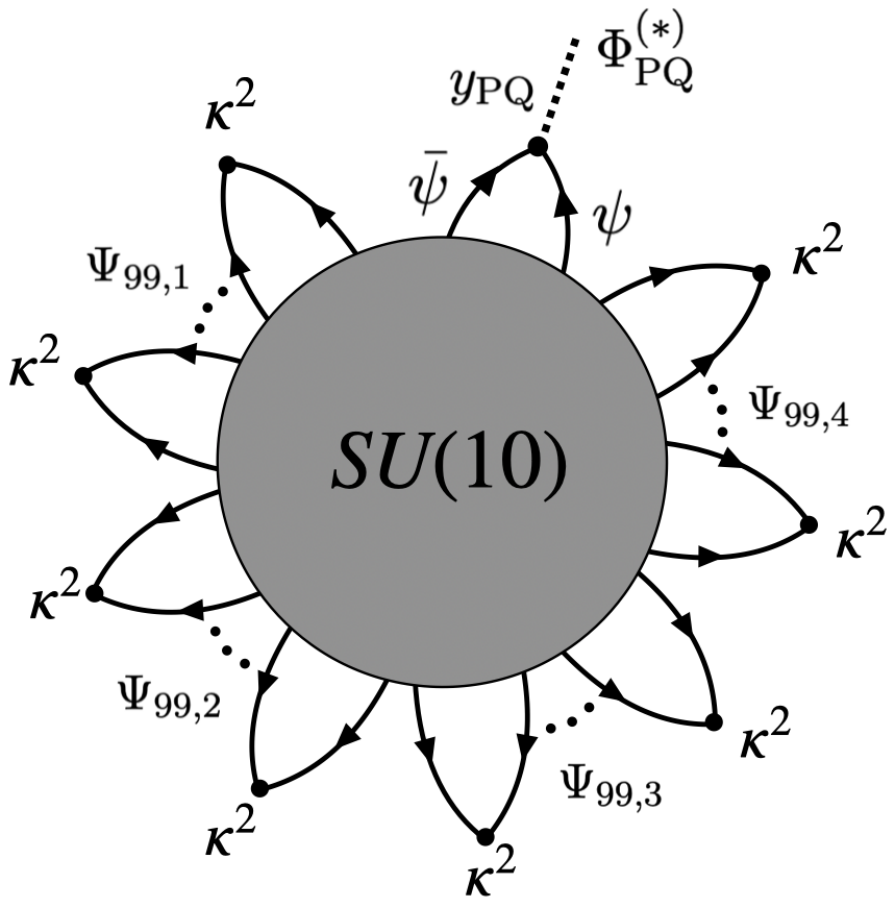} 
    \caption{'t Hooft vertex for the $SU(10)$ instanton effect. Each flavor of $\Psi_{99,i}$ has $2T({\rm Adj})=20$ legs,
    which are closed by 10 mass vertices. In the figure, $M=1$ for simplicity.}
    \label{fig:tHooft}  
\end{figure}

Let us estimate a small instanton effect of the $SU(10)$ sector.
The 't Hooft vertex is illustrated in Fig.~\ref{fig:tHooft}.
Using Instanton NDA presented in ref.~\cite{Csaki:2023ziz}, our estimation of the small instanton effect
for the axion potential is
\begin{align}
&V_{\rm bias} \approx C_{10}\left(\frac{2\pi}{\alpha_{\rm UV}(M)}\right)^{2\times 10}(\Phi_{\rm PQ}+\Phi_{\rm PQ}^*)\notag\\&\times\int\frac{d\rho}{\rho^5}\left(\Lambda_{SU(10)}\rho\right)^{b_0}e^{-2\pi^2\rho^2 M^2}
y_{\rm PQ}(\kappa^2 M\rho)^{10N_F}\rho \nonumber \\
 &\approx(\kappa^2)^{10N_F} C_{10} \left(\frac{2\pi}{\alpha_{\rm UV}(M)}\right)^{2\times 10} e^{-2\pi^2}\notag\\
 &~~~~~~~~~~~~~~~~~~~~~~\times\frac{\Phi_{\rm PQ}}{M} M^4 e^{-2\pi/\alpha_{\rm UV}(M)}+c.c. \, ,
\end{align}
where we take $1/\alpha_{\rm UV}(M) \equiv 4\pi / g_{\rm UV}^2 (M)$ as $1/3$ in the following estimation, $\Lambda^{b_0}_{SU(10)}=M^{b_0} e^{-\frac{8\pi^2}{g_{\rm UV}^2(M)}}$,
$b_0$ denotes the one-loop beta function coefficient, and $y_{\rm PQ}$ is the Yukawa coupling of $\Phi_{\rm PQ}$
and KSVZ fermions. 
The suppression factor of $e^{-2\pi^2\rho^2 M^2}$ originates from the breaking of $SU(10)$. 
The coefficient $C_{10}$ is the $SU(10)$ instanton density, defined as~\cite{Csaki:2019vte,Csaki:2023ziz}
\begin{align}
    C_N=\frac{K_1 e^{-(S^{(1/2)}-F^{(1/2)})\alpha(1/2)-(S^{(1)}-F^{(1)})\alpha(1)}}{(N-1)!(N-2)!}e^{-K_2 N}\, .
\end{align}
Here, $N=10$, $K_1\approx0.466$, $K_2\approx 1.678$, $\alpha(1/2)=0.145873$, $\alpha(1)=0.443307$.
$S^{(t)}$ and $F^{(t)}$ are respectively the numbers of scalar and fermion multiplets with isospin $t$ under $SU(2)$ of the $SU(10)$ instanton.
We then utilize the following estimation in our analysis,
\begin{align}
\label{eq:bias}
  V_{\rm bias}&= 3\times 10^2 (\kappa^2)^{40} \frac{e^{-2\pi/\alpha_{\rm UV}}}{\alpha_{\rm UV}^{20}} e^{-2\pi^2} M^3 \Phi_{\rm PQ}+c.c.\notag\\
   &=6\times 10^2  \epsilon \frac{e^{-2\pi/\alpha_{\rm UV}}}{\alpha_{\rm UV}^{20}} e^{-2\pi^2} M^3 v_{\rm PQ}\cos(\theta_a
   )\ .
\end{align}
This potential provides a bias term in the axion potential because it corresponds to the case with the domain wall number $N_{\rm DW} = 1$
(see Eq.~\eqref{QCDeffect}),
and $\epsilon\equiv(\kappa^2)^{40}$ is introduced to evaluate the bias term for later convenience.
Small instanton effects from $SU(5)_V$, $SU(3)_C$, and $SU(2)_L$ are negligibly small due to the exponential suppression by small gauge couplings, the flavor structure of the SM quarks, $\kappa$, and light $SU(3)_C$ octet and $SU(2)_L$ triplet fermions.

\section{Post-inflationary axion}\label{cosmology_section}

Let us consider the scenario that the $U(1)_{\rm PQ}$ symmetry is spontaneously broken by the VEV of $\Phi_{\rm PQ}$ after reheating. At a temperature around $v_{\rm PQ}$, the spontaneous breaking of the PQ symmetry gives rise to cosmic strings, corresponding to a winding of $\theta_a:0\to 2\pi$.
We focus on the scenario in which the axion field starts to oscillate due to the axion mass originated from the bias term, $m_{\rm bias}^2 \equiv \frac{\partial^2 V_{\rm bias}/\partial \theta_a^2}{v_{\rm PQ}^2}$, before non-perturbative QCD effects become effective, 
\begin{align}
m_a(T_1)\approx m_{\rm bias} = 3H(T_1)\ ,
\end{align}
where $m_a(T)$ denotes the axion mass at temperature $T$, $T_1$ represents the temperature when the oscillation starts, and $H(T_1)$ is the Hubble parameter,
\begin{align}
    H(T_1)^2\approx \frac{\pi^2}{90M_{\rm Pl}^2}g_*T_1^4\ .
\end{align}
Here, $g_*$ is the number of effective relativistic degrees of freedom of the energy density.
We require
\begin{align}
\label{eq:t1_qcd}
    T_1>0.98 \, {\rm GeV}\left(\frac{v_{\rm PQ}/3}{10^{12} \, {\rm GeV}}\right)^{-0.19}\equiv T_{1,\rm QCD}\ ,
\end{align}
where the right-hand side denotes the temperature when the axion would start to oscillate with non-perturbative QCD effects if there was no bias term.
Below $\sim T_1$, a string is attached by a domain wall, and the domain wall tension can collapse the string-wall system. This scenario is similar to the standard $N_{\rm DW}=1$ case, while the domain wall can be formed before the QCD transition temperature in the present scenario.
The axion abundance is estimated in a similar way as the $N_{\rm DW}=1$ case, except that the bias term contribution is time-independent,
\begin{align}
 \Omega_ah^2 \approx   2\times 10^{-12}  \, \frac{v_{\rm PQ}}{T_1}\ .
\end{align}
However, a precise determination of the abundance remains uncertain and requires dedicated numerical simulations.

A large bias term shifts the potential minimum of the QCD axion potential originated from $V_{\rm QCD}$,
and for the estimation of the amount of the shift, we use
\begin{align}
    {\mit \Delta} \bar\theta\equiv \frac{m_{\rm bias}^2v_{\rm PQ}^2}{m_a^2 F_a^2} \bar\theta_c \ ,
\end{align}
where $m_a$ represents the axion mass from the total axion potential. We focus on $m_{\rm bias}\ll m_a$ and require
\begin{align}
     {\mit \Delta}\bar\theta\lesssim 10^{-10}\ ,
\end{align}
to satisfy the experimental upper bound of the neutron EDM~\cite{Abel:2020pzs}.

Since our model contains exotic particles, we need to discuss their cosmological implications. 
After the breaking into $SU(5)_V$, the KSVZ fermions transform as ${\bf 10}$ and $\overline{\bf 10}$ under $SU(5)_V$, and can mix with the SM sector via
\begin{align}
    \mathcal{L}\ni \sum_f\Phi_a^{ij} \bar\psi^a {\bf 10}^{(1)}_{ij, f} \ ,
\end{align}
allowing them to decay into SM particles.
Additionally, the theory contains $SU(2)_L$ triplet fermions and $SU(3)_C$ octet fermions.
They remain in thermal equilibrium with the SM sector when the reheating temperature is higher than their masses.
To facilitate their decay, we assume the triplet couples to the Higgs and left-handed lepton via an interaction of the form $\mathcal{L}\sim \widetilde W H L $ where $\widetilde W,H,L$ denote the triplet, SM Higgs, and left-handed lepton. We also assume that the octet couples to the color triplet Higgs and down-type quark via $\mathcal{L}\sim \widetilde G H_C^\dagger \bar d$ where $\widetilde G,H_C,L$ denote the octet, color triplet Higgs, and right-handed down-type quark, respectively. 
The light color triplet Higgs may induce a rapid proton decay (see e.g. ref.~\cite{Nath:2006ut} for the estimation of the decay rate), which is strongly constrained by Super-Kamiokande~\cite{Super-Kamiokande:2014otb,Super-Kamiokande:2020wjk} and further explored in future experiments~\cite{Hyper-Kamiokande:2018ofw,DUNE:2020ypp,JUNO:2022qgr}. To satisfy current bounds, the triplet Higgs mass is assumed to be greater than about $5\times 10^{12}$\,GeV.
Through the assumed interactions, $SU(2)_L$ triplet fermions and $SU(3)_C$ octet fermions safely decay into the SM sector before they dominate the Universe.

\begin{figure}[!t]
    \centering
    \includegraphics[width=0.4\textwidth]{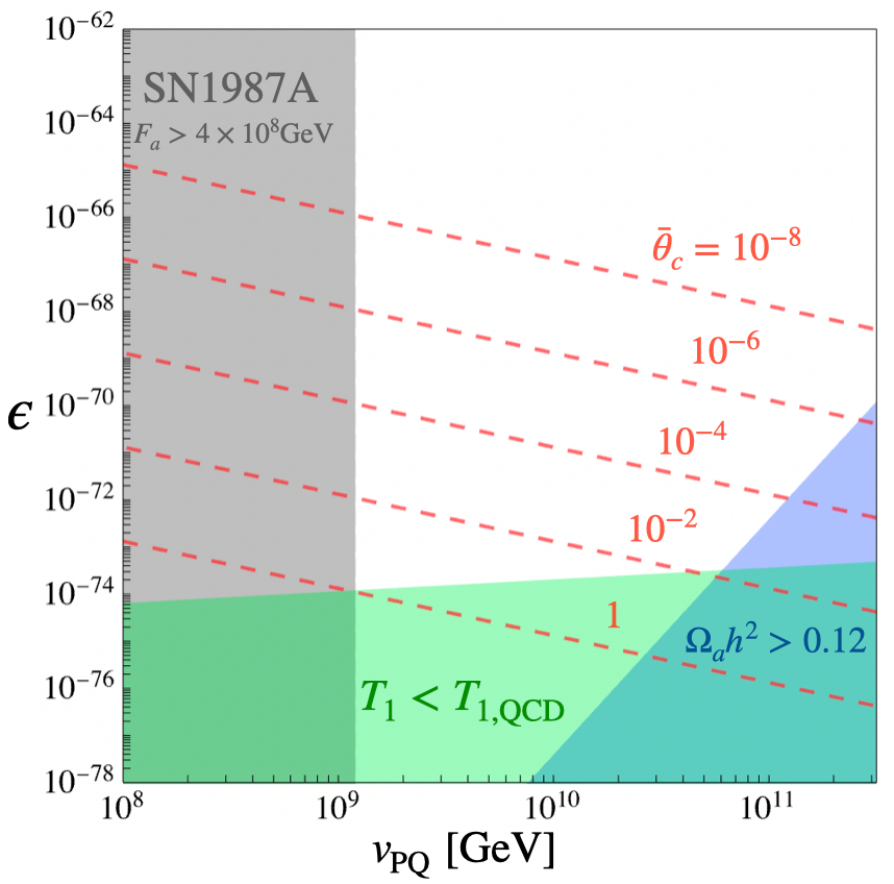} 
    \caption{Constraints on $v_{\rm PQ}-\epsilon$ plane. The gray-shaded region denotes the constraint from $F_a>4\times 10^8$\,GeV (SN1987A). The green-shaded region corresponds to $T_1<T_{1,\rm QCD}$.
The blue-shaded region represents $\Omega_{a}h^2>0.12$,
and its boundary gives the correct axion DM abundance. 
The red-dashed lines indicate contours of ${\mit \Delta}\bar\theta=10^{-10}$ for different values of $\bar\theta_c$.}
    \label{fig:constraint}  
\end{figure}

Fig.~\ref{fig:constraint} summarizes constraints on $v_{\rm PQ}-\epsilon$ plane. The gray-shaded region denotes the constraint from $F_a>4\times 10^8$\,GeV (SN1987A)~\cite{Carenza:2019pxu}, while the green-shaded region corresponds to $T_1<T_{1,\rm QCD}$.
The blue-shaded region represents $\Omega_{a}h^2>0.12$,
and its boundary gives the correct axion DM abundance. 
The red-dashed lines indicate contours of ${\mit \Delta}\bar\theta=10^{-10}$ for different values of $\bar\theta_c$,
which is naturally smaller than $\sim 10^{-2}$ in our model.
The correct axion DM abundance predicts $v_{\rm PQ} \gtrsim 6 \times 10^{10} \, \rm GeV$.

\section{Conclusions and discussions
\label{sec:Discussion}}

We have presented a new GUT model equipped with a viable post-inflationary QCD axion.
Our GUT is based on a symmetry breaking pattern,
$SU(10) \times SU(5)_1 \rightarrow SU(5)_V \supset SU(3)_C \times SU(2)_L \times U(1)_Y$,
where $SU(5)_1$ and a special embedding of $SU(5)_2\subset SU(10)$ are broken to a diagonal subgroup $SU(5)_V$.
The model includes a vector-like pair of KSVZ fermions which transform as the (anti-)fundamental representations of $SU(10)$, ensuring that the domain wall number is \( N_{\text{DW}} = 1 \). However, after the GUT breaking, the number of KSVZ quark pairs exceeds one, which apparently suggests a domain wall problem.
The inconsistency is resolved by the presence of small instanton effects, which generate a PQ-violating bias term in the axion potential. This bias term explicitly lifts the vacuum degeneracy and enables the decay of the string-wall network.
In our model, the correct axion DM abundance predicts $v_{\rm PQ} \gtrsim 6 \times 10^{10} \, \rm GeV$, a regime that is being explored by ongoing and future experiments, some of which cover or extend into this range~\cite{AxionLimits,Baryakhtar:2018doz,BREAD:2021tpx,Aja:2022csb,McAllister:2017lkb,Quiskamp:2024oet,Beurthey:2020yuq,Garcia:2024xzc,Fan:2024mhm,Lawson:2019brd,DeMiguel:2023nmz,QUAX:2024fut,Ahyoune:2023gfw,DMRadio:2022pkf,Crisosto:2019fcj,ADMX:2018gho,ADMX:2020ote,ADMX:2021nhd,ADMX:2024xbv}.

To demonstrate our framework, we have focused on a non-supersymmetric GUT model,
where a precise gauge coupling unification is not automatic.
However, there is no apparent obstacle to consider a supersymmetric version of the framework,
which, as is well known, gives an automatic gauge coupling unification.
It is interesting to note that gauginos play a role in suppressing small instanton effects on the axion potential
as was done by extra fermions in the non-supersymmetric model.
Therefore, to consider a supersymmetric model and estimate small instanton effects is a natural future direction.
Since a supersymmetric axion model generally predicts the saxion and axino,
the study of their cosmological implications is important. 

To successfully resolve the strong CP problem, the PQ symmetry must be preserved to an exceptionally high degree of accuracy, while it has been widely argued that global symmetries are generically violated by quantum gravitational effects.
A promising direction to address this so-called axion quality problem is to consider extra dimensional theories.
For instance, refs.~\cite{Choi:2003wr,Reece:2024wrn}
identify the axion as the fifth component of a five-dimensional gauge field,
where the PQ symmetry is not imposed explicitly, but rather emerges accidentally due to higher-dimensional gauge invariance.
Alternatively, one can place the axion in a warped extra dimension. The axion is localized in the bulk
and far from a boundary where PQ-violating effects originate
\cite{Flacke:2006ad,Cox:2019rro,Bonnefoy:2020llz,Lee:2021slp}. According to the AdS/CFT correspondence, such warped axion models can be interpreted as the holographic duals of four-dimensional conformal axion models
\cite{Nakai:2021nyf,Nakagawa:2023shi,Nakagawa:2024kcb}.

Dimensional deconstruction~\cite{Arkani-Hamed:2001kyx} suggests that a five-dimensional gauge theory is reproduced in a four dimensional theory by replacing the continuous extra dimension with a chain of gauge groups linked by scalar fields. A gauge field propagating in the extra dimension manifests as a Kaluza-Klein (KK) tower in the 5D theory, while in the deconstructed 4D theory, the scalar link fields acquire VEVs, breaking the extended gauge symmetry down to a diagonal subgroup and generating a discrete KK-like mass spectrum.
Our product GUT model was inspired by this dimensional deconstruction. Therefore, it is natural to embed our framework into
a higher dimensional theory to address the axion quality problem,
where the bulk of the compact extra dimension respects a $SU(10)$ gauge symmetry, while one of the boundaries explicitly breaks the $SU(10)$ to the special $SU(5)$ subgroup. All the SM matter fields are localized on this boundary and transform under the $SU(5)$. CP symmetry is also spontaneously broken at this boundary to generate the CKM phase.

To incorporate a high-quality axion into the 5D setup, we may consider the axion as the fifth component of a 5D gauge field,
or the axion localized in a warped extra dimension.
The size of small instanton effects is expected to be different in two cases, and its calculation is essential to clarify the model predictions. 
We leave it to a future exploration.

\section*{Acknowledgments}

We would like to thank Shota Nakagawa for helpful discussions. Y.N. is supported by Natural Science Foundation of Shanghai. M.S. is supported by the MUR projects 2017L5W2PT.
M.S. also acknowledges the European Union - NextGenerationEU, in the framework of the PRIN Project “Charting unexplored avenues in Dark Matter” (20224JR28W).

\newpage

\appendix
\begin{widetext}  
\section{$SU(5)\subset SU(10)$}\label{app:SU(5)}
We consider a special embedding of $SU(5)$, 
\begin{align}
    SU(5) \subset SU(10).
\end{align}
In this embedding, the $\mathbf{10}$ representation of $SU(5)$ is embedded in the fundamental representation $\mathbf{10}$ of $SU(10)$, which is the second-order anti-symmetric tensor $\Psi_{ij}$ ($i,j = 1,\dots,5$):
\begin{align}
    \Psi_{ij}=-\Psi_{ji}.
\end{align}
Explicitly, the $5\times 5$ matrix representation of $\Psi$ is given by
\begin{align}
\label{eq:psi_matrix}
  \Psi=
 \begin{pmatrix}
  0 & A & C & D & G\\
  -A & 0 & B & E & H\\
  -C & -B & 0  & F & I\\
  -D &  -E & -F & 0 & J\\
  -G & -H & -I &-J & 0
 \end{pmatrix}.
 \end{align}
Under an infinitesimal $SU(5)$ transformation,
\begin{align}
   \Psi_{ij} \rightarrow  U_{i\bar i}\Psi_{\bar i\bar j} (U^T)_{\bar j j}
    = U_{i\bar i} U_{j \bar j} \Psi_{\bar i\bar j}
    \simeq \left(\delta_{i\bar i}\delta_{j\bar j}+
    (i\alpha^a T^a)_{i\bar i}\delta_{j\bar j}+
   \delta_{i\bar i} (i\alpha^a T^a)_{j\bar j} \right)\Psi_{\bar i\bar j},
\end{align}
where $U=\exp(i\alpha^a T^a)$ with $T^a$ ($a=1,\dots,24$) being the generators of $SU(5)$ and $\alpha^a$ are real parameters. The last approximation holds for small $\alpha^a$.
The transformation of each element in Eq.~\eqref{eq:psi_matrix} is given by
\begin{equation}
\begin{split}
\label{eq:trans_10}
&-i\, A\to -B \alpha T_{13} - E \alpha T_{14} - H \alpha T_{15} + A (1 + \alpha T_{11} + \alpha T_{22}) + C \alpha T_{23} + D \alpha T_{24} + G \alpha T_{25}, \\[1.5ex]
&-i\, B\to   C \alpha T_{21} - F \alpha T_{24} - I \alpha T_{25} - A \alpha T_{31} + B (1 + \alpha T_{22} + \alpha T_{33}) + E \alpha T_{34} + H \alpha T_{35}, \\[1.5ex]
&-i\,C\to B \alpha T_{12} - F \alpha T_{14} - I \alpha T_{15} + A \alpha T_{32} + C (1 + \alpha T_{11} + \alpha T_{33}) + D \alpha T_{34} + G \alpha T_{35}, \\[1.5ex]
&-i\,D\to  E \alpha T_{12} + F \alpha T_{13} - J \alpha T_{15} + A \alpha T_{42} + C \alpha T_{43} + D (1 + \alpha T_{11} + \alpha T_{44}) + G \alpha T_{45}, \\[1.5ex]
&-i\,E\to  D \alpha T_{21} + F \alpha T_{23} - J \alpha T_{25} - A \alpha T_{41} + B \alpha T_{43} + E (1 + \alpha T_{22} + \alpha T_{44}) + H \alpha T_{45}, \\[1.5ex]
&-i\,F\to   D \alpha T_{31} + E \alpha T_{32} - J \alpha T_{35} - C \alpha T_{41} - B \alpha T_{42} + F (1 + \alpha T_{33} + \alpha T_{44}) + I \alpha T_{45}, \\[1.5ex]
&-i\,G\to  H \alpha T_{12} + I \alpha T_{13} + J \alpha T_{14} + A \alpha T_{52} + C \alpha T_{53} + D \alpha T_{54} + G (1 + \alpha T_{11} + \alpha T_{55}), \\[1.5ex]
&-i\,H\to  G \alpha T_{21} + I \alpha T_{23} + J \alpha T_{24} - A \alpha T_{51} + B \alpha T_{53} + E \alpha T_{54} + H (1 + \alpha T_{22} + \alpha T_{55}), \\[1.5ex]
&-i\,I\to  G \alpha T_{31} + H \alpha T_{32} + J \alpha T_{34} - C \alpha T_{51} - B \alpha T_{52} + F \alpha T_{54} + I (1 + \alpha T_{33} + \alpha T_{55}), \\[1.5ex]
&-i\,J\to   G \alpha T_{41} + H \alpha T_{42} + I \alpha T_{43} - D \alpha T_{51} - E \alpha T_{52} - F \alpha T_{53} + J (1 + \alpha T_{44} + \alpha T_{55}),
\end{split}
\end{equation}
where we have omitted the index $a$ for $\alpha^a T^a$ for notational simplicity.

We now rewrite the $5\times 5$ matrix $\Psi$ as a ten-dimensional column vector:
\begin{align}
    \Psi =
    \begin{pmatrix}
    A\\
    B\\
    C\\
    D\\
    E\\
    F\\
    G\\
    H\\
    I\\
    J
    \end{pmatrix}.
\end{align}
Under an infinitesimal $SU(5)$ transformation, $\Psi$ transforms as
\begin{align}
    \Psi \rightarrow \left( 1+i\, \mathcal{H} \right) \Psi.
\end{align}
Here, the generator matrix $\mathcal{H}$ is given by
\begin{align}
&\mathcal{H}=\\
&\tiny  \left(
\begin{array}{cccccccccc}
 \alpha T_{11} + \alpha T_{22}  & -\alpha T_{13} & \alpha T_{23} & \alpha T_{24} & -\alpha T_{14} & 0
   & \alpha T_{25} & -\alpha T_{15} & 0 & 0 \\
 -\alpha T_{31} & \alpha T_{22} + \alpha T_{33}  & \alpha T_{21} & 0 & \alpha T_{34} & -\alpha T_{24}
   & 0 & \alpha T_{35} & -\alpha T_{25} & 0 \\
 \alpha T_{32} & \alpha T_{12} & \alpha T_{11} + \alpha T_{33}  & \alpha T_{34} & 0 & -\alpha T_{14} &
   \alpha T_{35} & 0 & -\alpha T_{15} & 0 \\
 \alpha T_{42} & 0 & \alpha T_{43} & \alpha T_{11} + \alpha T_{44}  & \alpha T_{12} & \alpha T_{13} &
   \alpha T_{45} & 0 & 0 & -\alpha T_{15} \\
 -\alpha T_{41} & \alpha T_{43} & 0 & \alpha T_{21} & \alpha T_{22} + \alpha T_{44} & \alpha T_{23} &
   0 & \alpha T_{45} & 0 & -\alpha T_{25} \\
 0 & -\alpha T_{42} & -\alpha T_{41} & \alpha T_{31} & \alpha T_{32} & \alpha T_{33} + \alpha T_{44} 
   & 0 & 0 & \alpha T_{45} & -\alpha T_{35} \\
 \alpha T_{52} & 0 & \alpha T_{53} & \alpha T_{54} & 0 & 0 & \alpha T_{11} + \alpha T_{55}  & \alpha T_{12}
   & \alpha T_{13} & \alpha T_{14} \\
 -\alpha T_{51} & \alpha T_{53} & 0 & 0 & \alpha T_{54} & 0 & \alpha T_{21} & \alpha T_{22} + \alpha T_{55} 
   & \alpha T_{23} & \alpha T_{24} \\
 0 & -\alpha T_{52} & -\alpha T_{51} & 0 & 0 & \alpha T_{54} & \alpha T_{31} & \alpha T_{32} & \alpha T_{33} + \alpha T_{55}  & \alpha T_{34} \\
 0 & 0 & 0 & -\alpha T_{51} & -\alpha T_{52} & -\alpha T_{53} & \alpha T_{41} & \alpha T_{42} & \alpha T_{43} & \alpha T_{44} + \alpha T_{55} \\
\end{array}
\right) ,
\end{align}
which satisfies the conditions,
\begin{align}
    \mathcal{H} = \mathcal{H}^\dagger, \quad \mathrm{tr}(\mathcal{H}) = 0 \ .
\end{align}
This indicates that $\mathcal{H}$'s are given by linear combinations of the generators for the fundamental representation
$\bf 10$ of $SU(10)$, which explicitly shows the embedding of $SU(5)\subset SU(10)$.
In summary,
\begin{align}
\label{embeddingrelation}
   \Psi_{ij} \rightarrow  U_{i\bar i}\Psi_{\bar i\bar j} (U^T)_{\bar j j}
   \sim \tilde{U}_{mn} \Psi_{n}  \ ,
\end{align}
with $m,n = 1-10$ and $\tilde{U} \simeq 1+i\, \mathcal{H}$.

\section{Gauge Boson Masses in $SU(10)\times SU(5)_1\to SU(5)_V$}
\label{gauge boson}
Let us verify whether the gauge bosons acquire masses consistently after the spontaneous breaking of 
\begin{align}
    SU(10) \times SU(5)_1 \to SU(5)_V.
\end{align}
The covariant derivative of $\Phi$ is given by
\begin{align}
    D_\mu \Phi^j_a
    =
    \partial_\mu \Phi^j_a - i g (A_{\mu}^m T^m)^b_a \Phi^j_b
    + i \tilde{g} (\tilde{A}_{\mu}^l \tilde{T}^l)^j_i \Phi^i_a \ ,
\end{align}
where $T^m$'s are the generators for the fundamental representation of $SU(10)$,
$\tilde T^l$'s are the generators for the {\bf 10} representation of $SU(5)_1$
embedded into the fundamental representation of $SU(10)$, 
the indices $a,b$ correspond to those of $SU(10)$,
while the indices $i,j$ belong to $SU(5)_1$,
and $A_\mu$, $\tilde{A}_\mu$ represent their corresponding gauge fields.
The mass terms for the gauge fields arise from the kinetic term of $\Phi$
by substituting the VEV $\langle  \Phi \rangle \propto  \delta^j_a$, 
\begin{align}
    \propto \left( g A^m_\mu T^m - \tilde{g} \tilde{A}^l_\mu \tilde{T}^l \right)^2.
\end{align}
From this expression, we can analyze the gauge boson masses. It follows that, among the total 24+99 gauge bosons, only 24 modes remain massless. This result is consistent with the expected breaking pattern of the gauge symmetry.

\section{Effective Quark Mass Matrices and the CKM Phase}
\label{app:meff}
Let us discuss the effective quark mass matrices after integrating out the heavy KSVZ quarks. 
The interactions \eqref{nelsonbarr} generate up and down-type quark mass matrices \eqref{uptypemass} and \eqref{downtypemass}. To integrate out the heavy quarks in the up-type quark sector, we perform a (flavor) $SU(5)$ rotation as in ref.~\cite{Valenti:2021rdu},
\begin{align}
    &(q_{uf}\,\ U\,\, Q_u )\to  (q_{uf}\,\ U\,\, Q_u ) \, T_Q  \ ,\\[1ex]
    &\begin{pmatrix}
     \bar u_{f'}\\
     \bar U\\
     \bar Q_u
     \end{pmatrix}\to 
     T_U\begin{pmatrix}
     \bar u_{f'}\\
     \bar U\\
     \bar Q_u
     \end{pmatrix}\ ,
\end{align}
where $T_Q$ and $T_U$ denote the elements of the $SU(5)$,
\begin{align}
T_Q 
    & =
     \begin{pmatrix}
         {\bf 1}-\frac{A^* A^T}{|A|^2}\big(1-\frac{v_{\rm PQ}}{M}\big) & 0 & -\frac{A^*}{M}\\
         0 & 1 & 0\\ 
         \frac{A^T}{M} &  0 & \frac{v_{\rm PQ}}{M} \\
     \end{pmatrix}\ ,\\[1ex]
 T_U
    & =
     \begin{pmatrix}
         {\bf 1}-\frac{A A^\dagger}{|A|^2}\big(1-\frac{v_{\rm PQ}}{M}\big) &  \frac{A}{M} & 0\\
         -\frac{A^\dagger}{M} & \frac{v_{\rm PQ}}{M} & 0\\
         0 & 0 & 1
     \end{pmatrix}\ .
\end{align}
Here, $|A|^2=A^\dagger A$ and $M \equiv \sqrt{|A|^2+v_{\rm PQ}^2}$.
We have
\begin{align}
    \mathcal{M}_u\,T_U
    &=
    \begin{pmatrix}
        m_u \big({\bf 1}-\frac{AA^\dagger}{|A|^2}(1-\frac{v_{\rm PQ}}{M})\big) & m_u \frac{A}{M} & A^*\\
             0 & M & 0\\
        0 & 0 & v_{\rm PQ}
    \end{pmatrix}\ ,
\end{align}
and then find
\begin{align}
\hat{\mathcal{M}}_u=    T_Q\,\mathcal{M}_u\,T_U
    &=
    \begin{pmatrix}
      \big( {\bf 1}-\frac{A^* A^T}{|A|^2}(1-\frac{v_{\rm PQ}}{M})\big)  m_u \big({\bf 1}-\frac{AA^\dagger}{|A|^2}(1-\frac{v_{\rm PQ}}{M})\big) &  \big( {\bf 1}-\frac{A^* A^T}{|A|^2}(1-\frac{v_{\rm PQ}}{M})\big)m_u \frac{A}{M} & 0\\
             0 & M & 0\\
        \frac{A^T}{M} m_u \big({\bf 1}-\frac{AA^\dagger}{|A|^2}(1-\frac{v_{\rm PQ}}{M})\big) & \frac{A^T}{M}  m_u \frac{A}{M}  & M
    \end{pmatrix}\ .
\end{align}
In a similar manner, we obtain
\begin{align}
\hat{\mathcal{M}}_d= 
 \begin{pmatrix}
      \big( {\bf 1}-\frac{A^* A^T}{|A|^2}(1-\frac{v_{\rm PQ}}{M})\big)  m_d & 0\\
       \frac{A^T}{M}m_d & M
 \end{pmatrix}\ .
\end{align}
Utilizing the Schur complement, one finds
\begin{align}
    \det \hat{\mathcal{M}}_u
   & =M^2 \det
    \left(
       \left( {\bf 1}-\frac{A^* A^T}{|A|^2} \left(1-\frac{v_{\rm PQ}}{M} \right)\right)  m_u \left({\bf 1}-\frac{AA^\dagger}{|A|^2}\left(1-\frac{v_{\rm PQ}}{M} \right)\right) 
       \right)\ ,\\[1ex]
    \det \hat{\mathcal{M}}_d
    &=M \det
    \left(
        \left( {\bf 1}-\frac{A^* A^T}{|A|^2}\left(1-\frac{v_{\rm PQ}}{M} \right)\right)  m_d
       \right)\ ,
\end{align}
which indicate that the low-energy effective $3\times 3$ matrices after integrating out the heavy quarks are given as
\begin{align}
    \hat{\mathcal{M}}_{u, \rm eff}&=\left( {\bf 1}-\frac{A^* A^T}{|A|^2} \left(1-\frac{v_{\rm PQ}}{M} \right)\right)  m_u \left({\bf 1}-\frac{AA^\dagger}{|A|^2}\left(1-\frac{v_{\rm PQ}}{M} \right)\right)  \ , \label{effective_up_mass} \\[1ex]
    \hat{\mathcal{M}}_{d, \rm eff}&= \left( {\bf 1}-\frac{A^* A^T}{|A|^2}\left(1-\frac{v_{\rm PQ}}{M} \right)\right)  m_d\ .
    \label{effective_down_mass}
\end{align}

\end{widetext}

\bibliography{reference}

\end{document}